\documentclass[conference]{IEEEtran}
\IEEEoverridecommandlockouts
\usepackage{cite}
\usepackage{amsmath,amssymb,amsfonts}
\usepackage{algorithmic}
\usepackage{graphicx}
\usepackage{textcomp}
\usepackage{xcolor}
\usepackage{comment}
\usepackage{pgfplots}
\pgfplotsset{compat=1.8}
\usepackage{tikz}
\usepackage{multirow}
\usepackage{makecell}
\usetikzlibrary{arrows,backgrounds,positioning,calc}

\definecolor{bleudefrance}{rgb}{0.19, 0.55, 0.91}
\definecolor{bananayellow}{rgb}{1.0, 0.88, 0.21}
\definecolor{parisgreen}{rgb}{0.31, 0.78, 0.47}
\newcommand{\argmax}{\operatornamewithlimits{argmax}}

\def\BibTeX{{\rm B\kern-.05em{\sc i\kern-.025em b}\kern-.08em
    T\kern-.1667em\lower.7ex\hbox{E}\kern-.125emX}}
\begin{document}

\title{Zero-Shot Audio Classification using Image Embeddings\\
}

\author{\IEEEauthorblockN{Duygu Dogan\IEEEauthorrefmark{1},
Huang Xie\IEEEauthorrefmark{1},
Toni Heittola\IEEEauthorrefmark{1},
Tuomas Virtanen\IEEEauthorrefmark{1}}
\IEEEauthorblockA{\IEEEauthorrefmark{1}Unit of Computing Sciences,
Tampere University, Finland}}

\maketitle

\begin{abstract}
Supervised learning methods can solve the given problem in the presence of a large set of labeled data. However, the acquisition of a dataset covering all the target classes typically requires manual labeling which is expensive and time-consuming. Zero-shot learning models are capable of classifying the unseen concepts by utilizing their semantic information. The present study introduces image embeddings as side information on zero-shot audio classification by using a nonlinear acoustic-semantic projection. We extract the semantic image representations from the Open Images dataset and evaluate the performance of the models on an audio subset of AudioSet using semantic information in different domains; image, audio, and textual. We demonstrate that the image embeddings can be used as semantic information to perform zero-shot audio classification. The experimental results show that the image and textual embeddings display similar performance both individually and together. We additionally calculate the semantic acoustic embeddings from the test samples to provide an upper limit to the performance. The results show that the classification performance is highly sensitive to the semantic relation between test and training classes and textual and image embeddings can reach up to the semantic acoustic embeddings when the seen and unseen classes are semantically similar.  
\end{abstract}

\section{Introduction}
Supervised learning methods have enjoyed widespread adoption in various fields (e.g., computer vision, audio analysis). These methods mainly require a large amount of annotated data to robustly learn the concepts in focus. Acquisition of such dataset typically requires manual labeling which is expensive and time-consuming. Additionally, these methods have the capability to learn only the concepts defined in the development stage of the models. In many practical scenarios, however, the application should have the ability to learn unseen concepts, e.g., when it is challenging to collect a sufficient amount of instances belonging to a target class that is rare or changes over time.

In order to overcome the aforementioned problems, zero-shot learning (also referred to as zero-data learning) was proposed in  \cite{b1}. Zero-shot models aim to learn the concepts that have not been introduced during training by exploiting the semantic information about those classes. Based on the semantic information, both seen and unseen classes can be projected into the same representation space that is used to bridge between different domains. Consequently, the model acquires the ability to further mimic human cognition by transferring the knowledge between seen and unseen concepts. 

Recently, zero-shot learning has drawn increased attention in computer vision tasks, e.g., in \cite{b1.1, b1.2, b1.3, b1.4, b1.5}. Due to the lack of labeled instances belonging to the unseen zero-shot classes, side information about the unseen classes is required. Commonly used side information for zero-shot object recognition is reviewed in \cite{b1.6}. The authors classified the side information into two sets: semantic attributes and beyond. Semantic attributes refer to the intrinsic characteristics or properties of visual classes. Semantic information beyond attributes includes representations that are learned from the textual descriptions of the classes. 

Despite the increasing popularity of zero-shot learning in computer vision tasks, its applications and consequently the investigation of semantic embeddings in the audio content analysis have been fewer in comparison. Zero-shot audio classification based on class embeddings with a bilinear model was studied in \cite{b2, b3}, and the work was extended with a non-linear model in \cite{b4}. The proposed approaches were based on the compatibility between the audio and the semantic embeddings of the corresponding concepts where class labels of the audio samples were used as semantic information by employing word embedding models in \cite{b2} and sentence embedding models in \cite{b3}. Zero-shot audio classification using image embeddings as semantic information has not been studied so far. 

In this paper, we extend the previous work in \cite{b4} by introducing visual semantic information to the nonlinear acoustic-semantic projection model. Furthermore, we evaluate the performance of textual and visual embeddings both individually and together based on the classification accuracy. We compare our results with the model using acoustic embeddings as side information. In this work, we focus on the following questions: i) how does the performance of zero-shot audio classification change with different semantic spaces, ii) is the classification performance affected by the semantic relation between the training and test classes, and iii) if so, how does the performance change based on the hierarchical order of the semantically related classes. 

The remainder of this paper is organized as follows. First, Section \ref{sec2} introduces the selected zero-shot learning model for audio classification. Then, Section \ref{sec3} describes the dataset and acoustic and semantic embeddings. Next, Section \ref{sec4} explains the evaluation setup and reports the experimental results. Finally, Section \ref{sec5} concludes the paper with a discussion. 

\section{Zero-Shot Learning for Audio Classification} \label{sec2}
In this section, we formalize the zero-shot audio classification setting and then describe the selected methodology. 

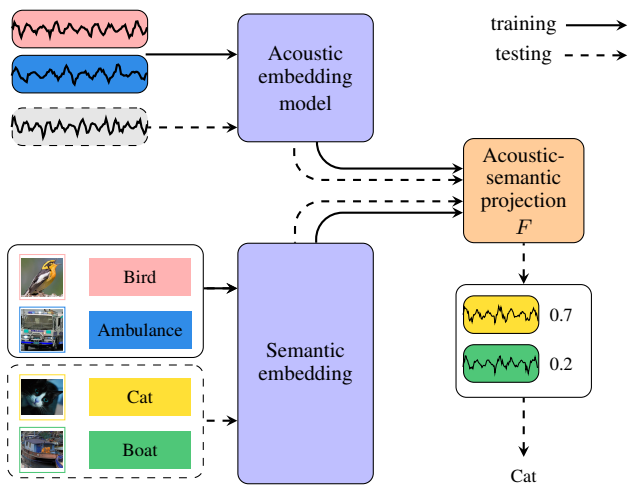
\begin{figure}[!ht]{}
	\centering
        \begin{tikzpicture}[
        declare function={
          excitation(\t,\w) = sin(\t*\w);
          noise = rnd - 0.5;
          source(\t) = excitation(\t,20) + noise;
          filter(\t) = 1 - abs(sin(mod(\t, 50)));
          speech(\t) = 1 + source(\t)*filter(\t);
        },
        scores_box/.style={rectangle, minimum width=2cm, minimum height=1.2cm,text centered, draw=black, fill=white},
        attn_square/.style={rectangle, minimum width=5cm, minimum height=1cm, minimum width=1cm, fill=red!30, draw=black},
        arrow1/.style={thick,<-,>=stealth, black!70},
        arrow2/.style={thick, dashed, <-,>=stealth, black!70},
        node distance=2cm]
        
        \node[minimum width=2cm, minimum height=2.1cm,text centered,yshift=-0.6cm] (audio_data) {};

        \node[rectangle, rounded corners, minimum width=1.8cm, minimum height=0.5cm,minimum width=1.8cm,text centered, draw=black, fill=red!30] (bird1) {};
        \draw let \p1=(bird1.west) in
            [black, thick, domain=0:1.8,samples=100,shift=(\p1)]
            plot ({\x},{(speech(100*\x)-1)/10});
            
        \node[rectangle, rounded corners, minimum width=1.8cm, minimum height=0.5cm,minimum width=1.8cm,text centered, draw=black, below of=bird1,yshift=1.4cm, fill=bleudefrance] (car1) {};
        \draw let \p1=(car1.west) in
            [black, thick, domain=0:1.8,samples=100,shift=(\p1)]
            plot ({\x},{(speech(80*(\x-9))-1)/10});
            
        \node[rectangle, text centered, below of=bird1,yshift=1.66cm, xshift=0.75cm] (audio_tr) {};
        
        \node[rectangle, dashed, rounded corners, minimum width=1.8cm, minimum height=0.5cm,text centered, draw=black, below of=car1,yshift=1.3cm, fill=gray!20] (x) {};
        \draw let \p1=(x.west) in
            [black, thick, domain=0:1.8,samples=100,shift=(\p1)]
            plot ({\x},{(speech(120*\x)-1)/10});
            
        \node[minimum width=2cm, minimum height=2.1cm,text centered, below of=audio_data] (image_data) {};
        
        \node[inner sep =0pt, rectangle, rounded corners, minimum width=2.6cm, draw=black, minimum height=1.5cm,text centered,below of=x, xshift=0.34cm, yshift=-0.32cm] (tr_semantic)
        {};
        
        \node[inner sep =0pt, rectangle, minimum width=0.6cm, draw=red!30, minimum height=0.6cm,text centered,below of=x, xshift=-0.5cm] (bird1_img)
        {\includegraphics[width=.03\textwidth]{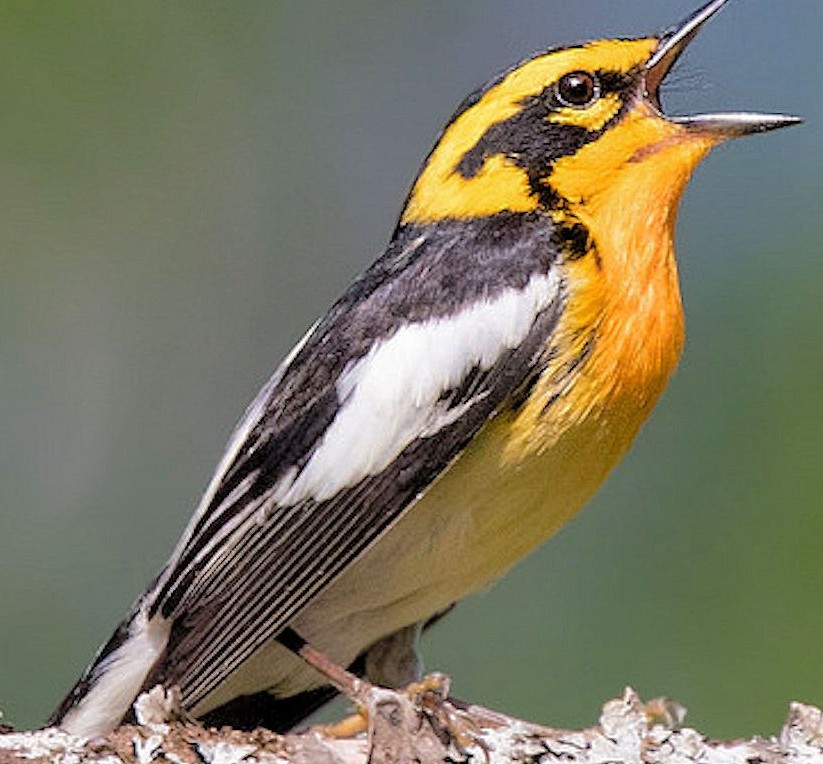}};
        
        \node[inner sep =0pt, minimum width=0.6cm, minimum height=0.6cm,text centered, below of=bird1_img, draw=bleudefrance, yshift=1.3cm] (car1_img)
        {\includegraphics[width=.03\textwidth]{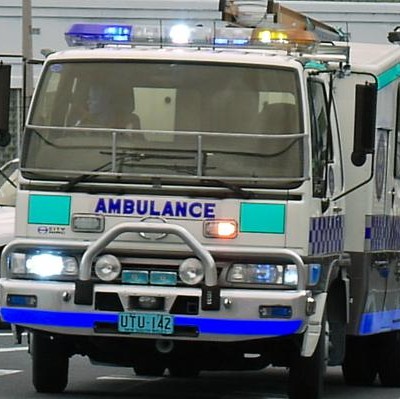}};
        
        \node[inner sep =0pt, rectangle, rounded corners, dashed, minimum width=2.6cm, draw=black, minimum height=1.5cm,text centered,below of=car1_img, xshift=0.84cm, yshift=0.78cm] (test_semantic)
        {};
        
        \node[inner sep =0pt, minimum width=0.6cm, minimum height=0.6cm,text centered, draw=bananayellow, below of=car1_img, yshift=1.1cm] (cat1_img)
        {\includegraphics[width=.03\textwidth]{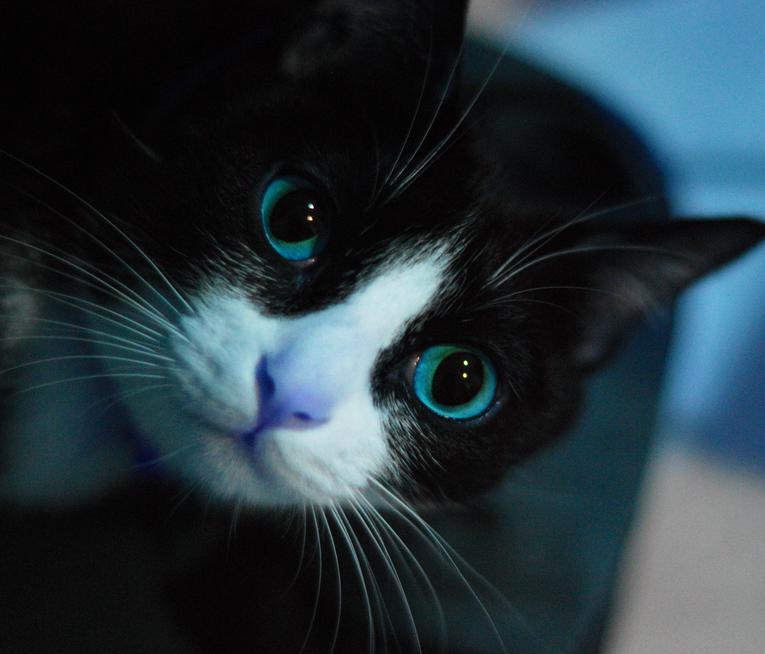}};
        
        \node[inner sep =0pt, minimum width=0.6cm, minimum height=0.6cm,text centered,draw=parisgreen, below of=cat1_img, ,yshift=1.3cm] (boat1_img)
        {\includegraphics[width=.03\textwidth]{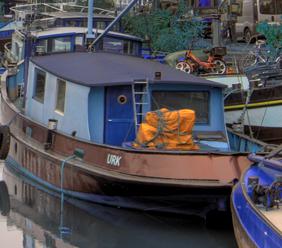}};
        
        \node[minimum width=2cm, minimum height=2.1cm,text centered, below of=image_data, xshift =0.5cm] (text_data) {};
        
        \node[minimum width=1.35cm, minimum height=0.5cm,text centered,right of=bird1_img, xshift=-0.7cm, fill=red!30] (bird_text)
        {\scriptsize\shortstack{Bird}};
        
        \node[minimum width=1.35cm, minimum height=0.5cm,text centered,right of=car1_img, xshift=-0.7cm,fill=bleudefrance] (car_text)
        {\scriptsize\shortstack{Ambulance}};
        
        \node[rectangle, fill=bananayellow, minimum width=1.35cm, minimum height=0.5cm,text centered,right of=cat1_img, xshift=-0.7cm] (cat_text)
        {\scriptsize\shortstack{Cat}};
        
        \node[rectangle, fill=parisgreen, minimum width=1.35cm, minimum height=0.5cm,text centered,right of=boat1_img, xshift=-0.7cm] (boat_text)
        {\scriptsize\shortstack{Boat}};
        
        \node[rectangle, rounded corners, minimum width=1.8cm, minimum height=3.2cm,text centered, draw=black, fill=blue!25, right of=car_text, xshift=0.2cm, yshift=-0.45cm] (hybrid_embed) {\footnotesize\shortstack{Semantic\\embedding}};
        
        \node[rectangle, rounded corners, minimum width=1.8cm, minimum height=1.7cm, yshift=1.8cm,text centered, draw=black, fill=blue!25, above of=hybrid_embed] (audio_embed) {\footnotesize\shortstack{Acoustic\\embedding\\model}};
        
        \node[rectangle, rounded corners, minimum width=1.6cm, minimum height=1cm,text centered, draw=black, fill=orange!40, right of=hybrid_embed, xshift=0.9cm, yshift=2.3cm] (F) {\footnotesize\shortstack{Acoustic-\\semantic\\projection\\ $F$}};
        
        \node[rectangle, rounded corners, minimum width=1.8cm, minimum height=1.5cm,text centered, draw=black, below of = F] (f) {};
        \draw [arrow1, black] (hybrid_embed.west) + (0,1cm) coordinate (s3) -- (tr_semantic.east |- s3);
        \draw [arrow1, black, dashed] (hybrid_embed.west) + (0,1cm) coordinate (s3) -- (test_semantic.east |- s3);
        \draw [arrow1, black,dashed] (hybrid_embed.220)  + (0,0cm) coordinate (s4) --  (test_semantic.0 |- s4);
        
        \draw [arrow1, black] (audio_embed.161) coordinate (s4) -- (audio_tr.0 |- s4);
        \draw [arrow1, black,dashed] (audio_embed.216) coordinate (s4) --  (x.east |- s4);
        
        \draw [arrow1, black, rounded corners=10pt] (F.160) coordinate (s3) -| (audio_embed.280);
        \draw [arrow1, black, dashed, rounded corners=10pt] (F.170) coordinate (s3) -| (audio_embed.260);
        
        \draw [arrow1, black, rounded corners=10pt] (F.200) coordinate (s3) -| (hybrid_embed.85);
        \draw [arrow1, black, dashed, rounded corners=10pt] (F.190) coordinate (s3) -| (hybrid_embed.95);
        
        \draw [arrow1, black, dashed] (f.north) coordinate (s3) -- (F.south);
        
        \node[rectangle, rounded corners, minimum width=1cm, minimum height=0.5cm,text centered, draw=black, below of=F,yshift=0.35cm, xshift=-0.3cm, fill=bananayellow] (l1) {};
        \draw let \p1=(l1.west) in
            [black, thin, domain=0:1,samples=100,shift=(\p1)]
            plot ({\x},{(speech(120*\x)-1)/10});
        \node[rectangle, rounded corners, minimum width=1cm, minimum height=0.5cm,text centered, right of=l1, xshift=-1.2cm] (l1) {\scriptsize 0.7};    
        \node[rectangle, rounded corners, minimum width=1cm, minimum height=0.5cm,text centered, draw=black, below of=F,yshift=-0.3cm, xshift=-0.3cm, fill=parisgreen] (l2) {};
        \draw let \p1=(l2.west) in
            [black, thin, domain=0:1,samples=100,shift=(\p1)]
            plot ({\x},{(speech(120*\x)-1)/10});
        \node[rectangle, rounded corners, minimum width=1cm, minimum height=0.5cm,text centered, right of=l2, xshift=-1.2cm] (l1) {\scriptsize 0.2};  
        
        \node[rectangle, minimum width=1.35cm, minimum height=0.5cm,text centered,below of=f, yshift=0.2cm] (label)
        {\scriptsize\shortstack{Cat}};

        \draw [arrow1, black, dashed] (label.north) coordinate (s3) -- (f.south);

        \node[above of = F, xshift=0cm,yshift=0.2cm] (tr0) {\footnotesize{training}};
        \node[right of = tr0,xshift=-0.5cm] (tr1) {};
        \node[below of = tr0, yshift=1.6cm] (test0) {\footnotesize{testing}};
        \node[right of = test0,xshift=-0.5cm] (test1) {};
        
        \draw [arrow1, black] (tr1.west) coordinate (s2) -- (tr0.east |- s2);
        \draw [arrow2, black] (test1.west) coordinate (s2) -- (test0.east |- s2);
        
        \end{tikzpicture}
    \caption[Zero-shot audio classification using visual semantic information.]{Zero-shot audio classification using visual semantic information. Training is represented in solid line and testing in dashed line. Sound classes \textit{bird} and \textit{ambulance} are used during training and \textit{cat} and \textit{boat} are used in testing.}
	\label{fig:pipeline}
\end{figure}

The pipeline of the zero-shot audio classification model is illustrated in Figure \ref{fig:pipeline}. First, audio feature embeddings are extracted from audio clips using an acoustic embedding model. Based on the semantic information that is used in zero-shot learning, semantic embeddings are obtained with an image embedding model that takes image samples or a language embedding model that takes textual class labels as input. Next, the model learns an acoustic-semantic projection by measuring the compatibility between two spaces during training. In the test time, the model takes unknown audio samples and applies the learned acoustic-semantic projection between the acoustic embedding of the unknown audio recordings and semantic embeddings of the new sound classes. Finally, a classifier selects the label whose semantic representation has the highest compatibility with the unknown audio sample.

We denote the audio sample space by $X$, and the semantic space of the seen and unseen sound classes by $Y$ and $Z$, respectively, where $Y\cap Z$ = $\emptyset$. Given an audio instance $x \in X$, and a sound class $y \in Y$, the main goal of the zero-shot learning is to learn a classifier $f$: $X \xrightarrow{} Y$, that is defined as 

\begin{equation}
    f(x) = \argmax_{y \in Y}F(x,y),
\end{equation}
where $F$ is the compatibility function between the representations of an audio instance and a sound class. 

In the present work, we project semantic embeddings onto acoustic embeddings due to the higher dimensionality of the semantic embeddings compared with acoustic embeddings. By doing so, we aim to capture the semantic embeddings in a lower and more relevant representational space. We define the compatibility function $F$ as the dot product between the acoustic embedding $\theta (x) \in R^{d_a}$ of an audio instance $x$ and the projection of the semantic embedding $\phi(y)\in R^{d_s}$ of a sound class $y$ as

\begin{equation}
    F(x,y) = H(\phi(y))^T\theta(x),
\end{equation}
where $H$ is the acoustic-semantic projection.

We follow the nonlinear acoustic-semantic projection model proposed for zero-shot audio classification in \cite{b4}. In order to decrease the computational complexity, the model uses factorization of a bilinear model and introduces nonlinear activation to capture the nonlinear relationship between acoustic and semantic embeddings. Consequently, we define $H$ as
\begin{equation}
    H(\phi(y))  = V^Tt(U^T\phi(y)),
    \label{eq1}
\end{equation}
where $U_{d_s \times r}$ and $V_{r \times d_a}$ are the learned projection matrices and $t$ is a nonlinear activation function.

The underlying assumption of zero-shot learning is that the acoustic embedding of an audio instance is expected to be closer to the  semantic embedding of the corresponding class in the acoustic-semantic space rather than those of other classes. Under this assumption, in test time, the model aims to assign an audio sample to an unseen class label that has the highest compatibility with.

\section{Audio, Visual, and Textual Embeddings} \label{sec3}
In this section, we describe the dataset and acoustic and semantic embeddings.
\subsection{Dataset}\label{sec3A}
To our knowledge, there is no joint audio and image dataset where mutual classes are defined. Therefore, we conducted our experiments based on separate datasets; the audio dataset, AudioSet \cite{b5} and image dataset, Open Images \cite{b6}. AudioSet is an unbalanced general audio dataset that contains 527 sound classes with over two million audio clips in total. The samples in AudioSet contain clip-level textual labels that consist of one or several words and a longer sentence description for every sound class. In this work, we use only the class labels rather than the sentence descriptions. Open Images is a dataset of around nine million images with image-level labels, object bounding boxes, object segmentation masks, and visual relationships. It contains a total of 16 M bounding boxes for 600 object classes on 1.9 M images. In our experiments, we use the images whose bounding boxes are defined. 

In order to semantically exploit every audio instance that belongs to one class by their visual representations, we defined mutual classes that occur in both AudioSet and Open Images. Both datasets provide the textual class labels with their Knowledge Graph Machine IDs (MID) \cite{b6-2} which are unique identifier codes for each entity in all languages. We define the mutual classes between AudioSet and Open Images by using the MID of the classes instead of string matching. Therefore, we prevent mistakenly capturing homonym word labels as well as losing information about the mutual classes that are defined with synonym words. Among 527 sound classes and 600 object classes, there are 79 mutual classes defined with the same MID in AudioSet and Open Images. 

AudioSet Ontology further provides a tree-like hierarchical structure for individual classes, where each audio sample is annotated with a more specific description when traversed to the deeper nodes. Using the hierarchical structure, we defined the child classes of each class until its deepest node. Among 79 mutual classes, we did not include the classes that have more than two children in the mutual class set (i.e. the class labels of Animal, Musical Instrument, Vehicle). In this way, we prevent the model from attempting to classify an audio instance either as "Animal" or "Cat". For the classes that have only one or two child classes in the mutual class set, we only kept the parent class in the mutual class set (i.e. Bird, Insect, Car, Aircraft, Tools) as the parent classes provide with more samples without significantly reducing the number of classes. We used the remaining 69 classes in our experiments. The least populated classes have 10 and 4 samples in AudioSet and Open Images, respectively. In order to provide a more balanced set, we randomly sampled the over-represented audio and image classes until each class has at most 300 and 1000 samples, respectively. 

\subsection{Acoustic Embeddings}\label{sec3B}
Following \cite{b2,b3,b4}, we used a pre-trained VGGish \cite{b7} to obtain acoustic embeddings. First, a 10-second audio clip was split into ten 1-second audio segments. Next, log mel spectrograms with the size of 96x64 were extracted for each segment and passed into VGGish that produced 128-dimensional audio embeddings. Then, the clip-level audio embeddings were obtained by averaging the embeddings extracted from the audio segments within the clip they belong to. It should be noted that the VGGish should be trained from scratch to reduce the bias arising from the seen classes used to train the model.

\subsection{Semantic Embeddings}\label{sec3C}
We compare class embeddings from three different domains: textual, audio, and image embeddings. We utilized pre-trained Word2Vec \cite{b8}, ResNet101 \cite{b9}, and VGGish \cite{b7} models for textual, image, and audio embeddings, respectively. Additionally, we concatenated the image and textual embeddings to capture the semantic information from both domains and denoted them as hybrid embeddings.

Word2Vec is a two-layered fully-connected neural network. The model learns the word embeddings by either predicting the current word by looking at its surroundings (bag-of-words) or the surrounding words from the current word (skip-gram). The model produces a 300-dimensional vector to represent the input word. We extracted the word embeddings based on the textual class labels of the sound classes defined in AudioSet. For the sound classes that are defined with multiple words, we averaged the extracted word embeddings.

ResNet101 is a 101-layered convolutional neural network that is trained on 1.28 million images composed of 1000 classes in the ImageNet database. The model takes images with the size of 224x224. First, we extracted the objects that belong to the mutual classes from the defined bounding boxes and resized them into 224x224. Then, we obtained a 1000-dimensional vector from the classification layer of the ResNet101 model for each image instance. We averaged the image embeddings over the samples within each class to represent the sound classes. It should be noted that the ResNet101 model with the current setup achieves an accuracy of 84\% on the supervised image classification task.

By the nature of zero-shot learning, the model aims to exploit semantic information assuming that there is not enough labeled data in test time. Therefore, in the task of zero-shot audio classification, it is not possible to use the audio data as semantic information. However, in this study, similar to the visual embeddings, we utilize the acoustic embeddings as semantic information by averaging the acoustic embeddings of audio instances that are used in the classification over the classes. As these semantic audio embeddings are calculated from the samples of test data, we use semantic audio embeddings with the purpose of measuring the known upper limit of the zero-shot learning performance that can be reached by using the semantic information provided with the test set itself.  

\section{Results and Analysis} \label{sec4}
In this section, we first describe the evaluation setup used to examine the effect of different semantic spaces and next present the experimental results of the zero-shot learning model. 
\subsection{Evaluation}
In our experiments, we evaluate zero-shot audio classification based on two different strategies, random-based class partitions, and category-based class partitions. On random-based class partition strategy, we randomly split the extracted dataset into three disjoint subsets for training, validation, and testing, which all at the end have 23 sound classes and a similar number of audio samples and image samples. We ran the experiments five times with different splits. Table \ref{tab1} displays the selected subsets on one trial with random-based class partitions. 

\begin{table}[!ht]
\caption{Selected subsets for one trial with the random-based class partitions}
\centering
\begin{tabular}{cc}
\textbf{Subset}        & \textbf{Class Labels}   \\ \hline \hline
\multicolumn{1}{c|}{Training} & \multicolumn{1}{l}{\begin{tabular}[l]{@{}l@{}}violin,
 bus,
 cat,
 door,
 horse,
 goose, 
 scissors, \\
 ratchet,
 sheep, 
 alarm clock,
 goat, 
 tap, 
 snake, \\
 chicken,
 car, 
 trombone,
 harp, 
 chainsaw,
 tool, \\
 frog,
 cupboard,
 coin,
 microwave oven\end{tabular}}                                            \\ \hline
\multicolumn{1}{c|}{Validation} & \multicolumn{1}{l}{\begin{tabular}[l]{@{}l@{}}camera,
 ambulance,
 motorcycle,
 aircraft,
 horn,
 duck, \\
 guitar, 
 drum,
 cattle, 
 mechanical fan,
 skateboard,
 chime, \\
 television, 
 bike,
 piano,
 harpsichord,
 sink,
 bird, \\
 bathtub,
 sewing machine,
 telephone,
 mouse,
 drawer\end{tabular}}                                                                  \\ \hline
\multicolumn{1}{c|}{Test} & \multicolumn{1}{l}{\begin{tabular}[l]{@{}l@{}}hand,
 boat,
 dog,
 computer keyboard,
 truck, 
 insect, \\
 flute,
 train,
 clock,
 pig,
 maracas, 
 banjo, 
 tableware, \\
 printer,
 trumpet,
 accordion, 
 saxophone, 
 cello, \\
 harmonica,
 blender,
 organ,
 hair dryer,
 toothbrush\end{tabular}}                                                \\ \hline \hline
\end{tabular}
\label{tab1}
\end{table}

On category-based class partition strategy, we conduct zero-shot learning within and across categories with the aim of investigating whether the hierarchical order of the semantically related classes affects the behavior of the classifier. We define a category as a class whose several child classes occur in the mutual classes, i.e. Animal, Musical Instrument, and Vehicle. Table \ref{tab2} shows the classes that belong to each of the categories. First, we conducted zero-shot learning within each category by randomly splitting the classes into training, validation, and test sets. In each split, the test sets of each category contain two different classes, and training and validation sets of the categories include 8, 6, and 4 different classes for Musical Instrument (MI), Animal (A), and Vehicle (V), respectively. Second, we evaluated the zero-shot learning across categories by combining two different categories and randomly splitting the classes where each subset has the same number of classes from both categories. When performed across categories, we included same number of classes from each category in all the subsets by limiting the number of classes based on the smallest category, i.e. Vehicle. We randomly selected four classes from each represented category for training and validation sets in each trial. In the test set, we combined one random class from each category. We ran the experiments five times with different splits for each strategy and category.
\vspace{-0.1cm}
\begin{table}[!ht]
\caption{Class labels for different categories}
\centering
\begin{tabular}{cc}
\textbf{Category}        & \textbf{Class Labels}  \\ \hline \hline
\multicolumn{1}{c|}{Animal } & \multicolumn{1}{l}{\begin{tabular}[l]{@{}l@{}}
cat,
horse,
goose, 
sheep, 
goat, \\
snake,
chicken, 
frog, 
duck, \\
cattle, 
bird, 
mouse,
dog,
pig \end{tabular}} \\ \hline
\multicolumn{1}{c|}{\begin{tabular}[c]{@{}c@{}}
Musical \\
Instrument \end{tabular}} & \multicolumn{1}{l}{\begin{tabular}[l]{@{}l@{}}
trombone,
harp, 
guitar, 
cello, 
drum, \\
organ, 
chime, 
piano,
harpsichord, 
flute,\\
maracas,  
banjo, 
trumpet,
accordion, \\ 
saxophone, 
harmonica,
horn,
violin\end{tabular}} \\ \hline 
\multicolumn{1}{c|}{Vehicle} & \multicolumn{1}{l}{\begin{tabular}[l]{@{}l@{}}
skateboard,
bus,
car, 
motorcycle, 
boat, \\
aircraft, 
bike, 
train,
truck, 
ambulance \end{tabular}} \\ \hline \hline
\end{tabular}
\label{tab2}
\end{table}

In all the setups, we implemented (\ref{eq1}) with two fully-connected layers where $r$ equals to $d_s$ and $t$ is the tanh activation function. We trained all the models using a batch size of 32, and SGD optimizer with the learning rate of $10^{-2}$ for 200 epochs. 

\subsection{Random-based zero-shot learning}
Random-based zero-shot audio classification results are reported in Table \ref{tab3}. The table shows the average and standard deviation of the accuracies with five trials where each trial has different randomization for class partitioning. In each run, training, validation, and test sets included 23 different classes. Overall, each model performs better than the random guess of 0.04. As the averaged audio embeddings represent the class-level audio instances that are used during classification, the model using audio embeddings determined the known upper limit for the performance. Between image and textual embeddings, we did not observe a significant advantage of choosing one over another in the current setup. Moreover, the model using the hybrid embeddings (image + textual) did not provide any improvement on the model performance. 
\begin{table}[ht!]
\caption{Zero-shot audio classification with random-based class partitions where random guess is 0.04}
\centering
\begin{tabular}{cc}
\multicolumn{1}{c}{\begin{tabular}[c]{@{}c@{}}
\textbf{Class} \\
\textbf{embeddings} \end{tabular}}        & \multicolumn{1}{c}{\begin{tabular}[c]{@{}c@{}}
\textbf{Accuracy [0--1]} \\
(mean $\pm$ SD)\end{tabular}}  \\ \hline \hline
\multicolumn{1}{c|}{Audio} & \multicolumn{1}{c}{\begin{tabular}[c]{@{}c@{}}
0.32 $\pm$ 0.02 \end{tabular}} \\ \hline \hline
\multicolumn{1}{c|}{Image} & \multicolumn{1}{c}{\begin{tabular}[c]{@{}c@{}}
\textbf{0.17 $\pm$ 0.02} \end{tabular}} \\ \hline
\multicolumn{1}{c|}{Text} & \multicolumn{1}{c}{\begin{tabular}[c]{@{}c@{}}
0.15 $\pm$ 0.04 \end{tabular}} \\ \hline
\multicolumn{1}{c|}{Hybrid} & \multicolumn{1}{c}{\begin{tabular}[c]{@{}c@{}}
0.16 $\pm$ 0.04 \end{tabular}} \\ \hline \hline
\end{tabular}
\label{tab3}
\end{table}

Additionally, we conducted experiments by randomly selecting audio and image embeddings among each class in each run to represent the class to which the selected instance belongs. The accuracy results of using audio and image embeddings with five runs were 0.16 $\pm$ 0.14 and 0.12 $\pm$ 0.03, respectively. Comparing the performance between averaging the embeddings over classes and randomly selecting, these results indicate that the averaging eliminates the undesirable effect of the outliers, and classes can be represented more accurately by taking advantage of the representational variety.

\subsection{Category-based zero-shot learning}
Zero-shot audio classification results within categories are reported in Table \ref{tab4}.  In all cases, the models perform better than the random guess of 0.50. The average of the accuracy scores shows that both individual and combined image and textual embeddings can easily reach up to the performance of audio embeddings when the seen and unseen concepts are semantically similar. Yet, the standard deviation of the accuracy scores indicates that the performance of the model is sensitive to the variation of the classes within the category. For instance, the accuracy of the model using hybrid embeddings is 0.75 when the test classes are selected as "boat" and "ambulance" whereas the accuracy is reported as 0.54 when the test set consists of "car" and "truck". In the hierarchical structure of the AudioSet, the class of "boat, water vehicle" is defined as the immediate subcategory of Vehicle whereas "car", "truck", and "ambulance" are defined under the "motor vehicle (road)". For further improvement, the intermediate categories should be explored. 

\begin{table}[ht!]
\caption{Top-1 accuracies (mean and SD) of zero-shot audio classification within categories where random guess is 0.50}
\centering
\begin{tabular}{cccc}

\multicolumn{1}{c||}{\multirow{2}{*}{\begin{tabular}[c]{@{}c@{}}
\\
\textbf{Class} \\
\textbf{Embedding} \end{tabular}}} & \multicolumn{3}{c}{\textbf{Categories}}   \\ \cline{2-4}    
\multicolumn{1}{c||}{}      & \multicolumn{1}{c}{Animal} & \multicolumn{1}{c}{\begin{tabular}[c]{@{}c@{}}
Musical \\
Instrument \end{tabular}} & \multicolumn{1}{c}{Vehicle}   \\ \hline \hline

\multicolumn{1}{c||}{Audio} & \multicolumn{1}{c}{0.61 $\pm$ 0.05} & \multicolumn{1}{c}{0.66 $\pm$ 0.09} & \multicolumn{1}{c}{0.61 $\pm$ 0.05}  \\ \hline \hline
\multicolumn{1}{c||}{Image} & \multicolumn{1}{c}{0.53 $\pm$ 0.07} & \multicolumn{1}{c}{0.57 $\pm$ 0.19} & \multicolumn{1}{c}{\textbf{0.62 $\pm$ 0.13}}  \\ \hline
\multicolumn{1}{c||}{Text} & \multicolumn{1}{c}{0.60 $\pm$ 0.06} & \multicolumn{1}{c}{0.60 $\pm$ 0.09} & \multicolumn{1}{c}{0.61 $\pm$ 0.10} \\ \hline \multicolumn{1}{c||}{Hybrid} & \multicolumn{1}{c}{\textbf{0.63 $\pm$ 0.08}} & \multicolumn{1}{c}{\textbf{0.62 $\pm$ 0.12}} & \multicolumn{1}{c}{0.59 $\pm$ 0.12} \\ \hline  \hline 
\end{tabular}
\label{tab4}
\end{table}

\begin{table}[ht!]
\caption{Top-1 accuracies (mean and SD) of zero-shot audio classification across categories where random guess is 0.50}
\centering
\begin{tabular}{cccc}

\multicolumn{1}{c||}{\multirow{2}{*}{\begin{tabular}[c]{@{}c@{}}
\textbf{Class} \\
\textbf{Embedding} \end{tabular}}} & \multicolumn{3}{c}{\textbf{Categories}}   \\ \cline{2-4}    
\multicolumn{1}{c||}{}      & \multicolumn{1}{c}{A + MI} & \multicolumn{1}{c}{A + V} & \multicolumn{1}{c}{MI + V}   \\ \hline \hline

\multicolumn{1}{c||}{Audio} & \multicolumn{1}{c}{0.92 $\pm$ 0.02} & \multicolumn{1}{c}{0.80 $\pm$ 0.18} & \multicolumn{1}{c}{0.94 $\pm$ 0.04}  \\ \hline \hline
\multicolumn{1}{c||}{Image} & \multicolumn{1}{c}{0.89 $\pm$ 0.03} & \multicolumn{1}{c}{0.77 $\pm$ 0.07} & \multicolumn{1}{c}{0.87 $\pm$ 0.15}  \\ \hline
\multicolumn{1}{c||}{Text} & \multicolumn{1}{c}{\textbf{0.90 $\pm$ 0.05}} & \multicolumn{1}{c}{0.84 $\pm$ 0.08} & \multicolumn{1}{c}{0.95 $\pm$ 0.04} \\ \hline \multicolumn{1}{c||}{Hybrid} & \multicolumn{1}{c}{0.88 $\pm$ 0.08} & \multicolumn{1}{c}{\textbf{0.85 $\pm$ 0.06}} & \multicolumn{1}{c}{\textbf{0.96 $\pm$ 0.04}} \\ \hline  \hline 
\end{tabular}
\label{tab5}
\end{table}

Zero-shot audio classification results across categories are reported in Table \ref{tab5}. As can be seen from the results, the model performs significantly better than the random guess of 0.50 in each setup. Compared with the results of within-category experiments, we observe that the model can easily learn different categories. This indicates that the classifier can transfer the knowledge from one class to another in the same category.

\section{Discussion and Conclusion} \label{sec5}
In this paper, we introduced image embeddings as semantic information for zero-shot audio classification. The zero-shot model was evaluated by comparing the performance of the models using visual, audio, and textual embeddings as semantic information. We used VGGish, ResNet101, and word2vec models to extract audio, image, and textual embeddings, respectively. We performed the audio classification task by extracting the mutually observed classes between audio and image datasets. 

The experiments showed that using visual semantic embeddings both individually and combined with textual embeddings can achieve better accuracy than a random guess on the zero-shot audio classification task. Yet, the difference between the results of using averaged and randomly selected semantic embeddings suggested that the performance of the model can be improved in the presence of a larger dataset where the averaging eliminates outliers more accurately. Additionally, the experimental results showed that the image, textual, and hybrid embeddings can further reach up to the performance of acoustic semantic embeddings when the seen and unseen classes are semantically similar. However, the varying behavior on the different splits indicates that the performance of the model is sensitive to the class splits when performed within the same category. The model reached up to the highest accuracy when performed across categories. 

Even though the visual embeddings might increase the representational power for each class by introducing rich variety to the model with a large number of image instances, they lack the ability to express more abstract classes. On the other hand, textual embeddings can capture the abstract classes in the presence of the nonambiguous textual class description. Yet, they provide each class only with a singular representation. Cognitively motivated approaches suggest that human semantic knowledge relies on perceptual information rather than only linguistic information \cite{b10}. In order to allow the model to capture complementary information from different modalities, future work should explore more advanced hybrid models. Furthermore, for a fully unbiased evaluation, the embedding models should be trained from scratch where the zero-shot classes are excluded from the training data.

\section*{Acknowledgment}
This research has been supported by the Academy of Finland, project no. 338502. The authors wish to acknowledge CSC-IT Center for Science, Finland, for computational resources used towards this research.

\vspace{12pt}
\end{document}